\documentclass[11pt,a4paper]{article}
\usepackage{amsfonts}
\usepackage{amssymb}
\usepackage{graphicx}
\usepackage{epsfig}
\usepackage{amsmath}
\usepackage{enumerate}
\RequirePackage{mathrsfs} \RequirePackage[sc]{mathpazo}
\RequirePackage{wasysym} \RequirePackage{setspace}

\makeatletter \@addtoreset{equation}{section}

\newcommand{\be}{\begin{equation}}
\newcommand{\ee}{\end{equation}}
\newcommand{\bea}{\begin{eqnarray}}
\newcommand{\eea}{\end{eqnarray}}
\begin{document}
\date{}
\title{   Lie symmetries and  2D   Material Physics}
\author{ Adil Belhaj$^{1}$ and Moulay Brahim Sedra$^{2}$
\hspace*{-8pt} \\
\\
{\small $^{1}$D\'epartement de Physique, Facult\'e
Polydisciplinaire, Universit\'e Sultan Moulay Slimane}\\{ \small
B\'eni Mellal, Morocco }
\\ {\small $^{2}$  LHESIR,   D\'{e}partement de Physique, Facult\'{e}
des Sciences, Universit\'{e} Ibn Tofail }\\{ \small K\'{e}nitra,
Morocco} } \maketitle

\begin{abstract}
Inspired from  Lie symmetry classification,  we establish   a
correspondence between rank two  Lie symmetries and 2D materials
physics. The material unit cell is
 accordingly interpreted as the  geometry of a  root system.  The
hexagonal   cells,  appearing  in  graphene like models,  are
analyzed in some details and  are  found to be associated with $A_2$
and $G_2$ Lie symmetries. This approach can be applied to  Lie
supersymmetries associated with  fermionic  degrees of freedom. It
has been suggested that  these extended symmetries  can   offer  a
new way to deal with  doping material geometries.  Motivated by Lie
symmetry applications in high energy physics, we speculate on a
possible connection with $(p,q)$ brane networks used in  the string
theory compactification on singular Calabi-Yau manifolds.
\end{abstract}

\newpage
\tableofcontents \thispagestyle{empty} \newpage \setcounter{page}{1}

\section{Introduction}
Several studies of  magnetic  physical properties of strongly
correlated electron models have been carried out,  in connection
with the elaboration of  nano-materials.  The most studied model is
the graphene using
  different calculation methods with appropriate approximations\cite{1,2}.
 The material is a monolayer of carbon crystal  forming
a two dimensional  hexagonal geometric lattice \cite{3}. Energy
spectrum of this material has a particular structure where the
valence and the conduction bands intersect at Dirac points producing
a semi-metal. In this way, low-energy excitations are described by a
pair of two-component fermions, equivalently, a four-component Dirac
fermion \cite{4}.

Recently, there have been many  attempts  to bluid   a bridge
between high energy physics and   the  graphene   using different
methods including
 the BTZ  black hole physics and the AdS/CFT correspondence  explored in
string theory  and related topics\cite{5,6}. More precisely,  a
stringy description in terms of  the (D3,D7) brane system  embedded
in type IIB superstring has been proposed in \cite{6}. In the
corresponding brane representation, each D3-D7 pair produces  a
complex massless two-component spinor living in the three
dimensional space-time. In connection with these activities,
  a brane realization of  the quantum Hall  effect,  based on  the  Calabi-Yau singularities classified by
    Lie symmetries,  has been  proposed
in \cite{7}. In particular, it has been suggested  a possible link
between  the  graphene   and a  special class of Lie symmetries
called indefinite.

More recently,  Lie symmetries has been  used  to investigate   a
class of materials engineered      from  the hexagonal structure. It
is recalled that this geometry  can be  considered as the most
stable one in nature which has been  explored  in many physical
applications, including   particle physics, string theory and
nano-technology. Indeed, an experimental treatment on the hexagonal
materials by the scanning tunneling spectroscopy (STS) has been
elaborated producing a $(\sqrt{3}\times \sqrt{3})R30^\circ$
supercell configuration  on the graphene and the silicene surfaces,
in contrast to the usual structure known by the (1$\times$1)
geometry\cite{8}. A close inspection in rank  two Lie symmetries has
showed   that the $(\sqrt{3}\times \sqrt{3})R30^\circ$  structure
appears naturally in the construction of the  root system of  the
$G_2$ exceptional Lie symmetry \cite{9,10}. Based on this
observation, a combination of $(\sqrt{3}\times \sqrt{3})R30^\circ$
and $(1\times 1)$ geometries has been developed  to engineer  new
materials relaying  on  a double hexagonal structure  arising  in
the root system of the $G_2$ Lie symmetry \cite{11}.

 Motivated by the above works,  we  establish  a  correspondence between 2D
material physics and    Lie symmetries.  In this way,  we interpret
the material  unit cells  as root systems of Lie  symmetries. This
may offer a new take on the geometrical elaboration of  2D
materials. Our focus is on rank two Lie symmetries. In particular,
we consider the hexagonal structure, and we find that it is linked
to $A_2$ and $G_2$ root systems. This method can be applied to  Lie
supersymmetries associated with fermionic degrees of freedom.  We
expect that  this class of symmetries  can offer  a new way to
approach doping material geometries.  Supported by the role placed
by Lie symmetries in high energy physics, we speculate on a possible
connection with $(p,q)$ brane webs used in the string theory
compactification on toric Calabi-Yau manifolds.

The paper is organized as follows. In section 2, we  give a short
overview on Lie symmetries.  Section 3 concerns a dictionary between
 the root systems of rank two  Lie symmetries and  2D  material physics.
Extension to Lie  supersymmetries  is  discussed in section 4.
Section 5 contains  concluding remarks and a  speculation from
string theory.

\section{  Lie symmetries}

We start by recalling that symmetry is one of the most important
ingredient  in  physics.  Precisely,  one remarks the crucial role
placed by Lie symmetries in   standard model  and higher dimensional
physical models  including superstrings, M and F-theories. In this
way,  the  root systems  have been explored to partially solve  many
problems arising  in such theories. A particular  emphasis put on
the hexagonal geometry   appearing in Lie symmetries used in string
theory compactification. It has been analyzed that many geometrical
background  relevant to particle physics and string theory  are
associated with
 the hexagonal symmetries  appearing in  the root systems \cite{12}.

The hexagonal structure arises also     in  lower dimensional
theories  including solid state physics dealing with the
graphene-like models. In the investigation of such materials, it has
been found the appearance of  new structures  shearing similarities
with the  hexagonal root systems of Lie symmetries \cite{8}.

Armed  by these works,  we establish a dictionary  between Lie
symmetries  and  2D material physics. Before going ahead, let us
note that the present study assumes some basic knowledge on Lie
symmetries and their root systems\cite{9,12}. More precisely, we
give a flash review on such mathematical backgrounds,   used in many
physical area including  high energy and condensed  matter physics.
Indeed, a Lie symmetry $g$ is a vector space together with an
antisymmetric bilinear bracket $[,] : g \times  g \to g$ satisfying
the Jacobi identity ($[a, [b,c]] + [c, [a, b]] + [b, [c,a]] = 0$).
It is realized  that any semi-simple Lie symmetry can be viewed as a
direct sum of simple Lie symmetries. The Cartan subalgebra $H$ is
generated by the all semi-simple elements, being the maximal abelian
Lie sub-algebra. It is observed  that $g$  may then be written as
the direct sum of $H$ and the subspaces $g_\alpha$:
\begin{equation}
 g=H\oplus \{\oplus \alpha g_\alpha\}
\end{equation}
where $\{g_\alpha=  x\in g|[ h,x ]=\alpha(x)x\}$ for $x\in g$. Here
$\alpha$  ranges over all elements of the dual of $H$. In Lie
theory, these vectors $\alpha$  are called roots.

Having discussed the Lie structure, we now  recall some basic
concepts of the  root systems  which will be explored  later on in
the discussion  of the desired correspondence.
Following \cite{9,10,12}, a root system $%
\Delta $ of a Lie symmetry   is defined as a subset of an  Euclidean
space $E$ satisfying the following constraints:
\begin{enumerate}
\item $\Delta $ is finite and spans $E$, $0$  is an element of $ \Delta $

\item if $\alpha$ is an element of  $ \Delta $,  then $k\alpha$ is also but  only for  $k=\pm 1$

\item  for any $ \alpha$ in $\Delta $,  $\Delta$ is invariant under reflection  $\sigma
_\alpha$, where $\sigma_{\alpha}(\beta)=\beta- 2\left\langle \beta
.\alpha \right\rangle \alpha $

\item if $\alpha $ et $\beta $ are two elements of $\Delta $, the quantity $
\left\langle \beta .\alpha \right\rangle =\frac{2(\beta .\alpha
)}{\left( \alpha .\alpha \right) }\in Z.$
\end{enumerate}
Note by the way that the root system $\Delta $ contains several
information  about the associated Lie symmetry structure. These
information will be relevant  in the present  discussion. In Lie
theory, it has been shown that there is a nice classification of
rank two Lie symmetries. Taking two root elements $\alpha $ and
$\beta $ of $\Delta $, it is evident to show the equation
\begin{equation}
\left\langle \beta ,\alpha \right\rangle \left\langle \alpha ,\beta
\right\rangle =4\cos ^{2}\theta
\end{equation}%
where $\theta $ is the angle between $\alpha $ and $\beta $. This
leads to the following constraint
\begin{equation}
0\leq \left\langle \beta ,\alpha \right\rangle \left\langle \alpha
,\beta \right\rangle \leq 4.
\end{equation}%
The possible values of $\theta $ are
$30^{0},45^{0},60^{0},90^{0},120^{0},135^{0}$ and $150^{0}$, leading
to a nice  classification.  This  generates four different Lie
symmetries  with the following dimensions
\begin{equation}
 \mbox{Dim}\;g = 2+|\Delta| \label{dim}
\end{equation}
where $|\Delta|$  denotes the number of the  roots associated with
$g$. The number  2 is called the rank identified  with  the number
of the simple roots,  being the dimension of the corresponding
Cartan subalgebra.  These  four symmetries, classified in terms of
the angle between simple roots $\theta$,  are listed  in  Table 1.
\begin{table}[h]
\begin{center}
\begin{tabular}{|c|c|c|}
  \hline
  Lie Symmetry & $\theta$ & $|\Delta|$ \\ \hline
   $A_1\bigoplus    A_1 $  & $90^{\circ }$& 4 \\ \hline
$A_2$ & $120^{\circ }$& 6  \\ \hline
   $B_2$ & $135^{\circ }$ & 8 \\ \hline
   $G_2$ & $150^{\circ }$& 12 \\
  \hline
\end{tabular}
\end{center}
\label{tab1} \caption{Classification of the  rank two  Lie
symmetries in terms of $\theta$ angle}
\end{table}
More details on this classification  and its physical applications
can be found in  literature.   An alternative way to classify these
symmetries  is to use  the Cartan matrices obtained from the scalar
product between  the  simple roots. These matrices $K = (k_{ij})$ of
size 2 take the following  general  form
\begin{equation} K=\langle\alpha_i, \alpha_j\rangle=\left(
  \begin{array}{cc}
    2 & k_{12} \\
    k_{21} & 2\\
  \end{array}
\right)
\end{equation}
where the values of   $k_{12}$ and $k_{21}$ are illustrated  in
table 2. It is recalled  that if the Cartan  matrix is symmetric,
the Lie symmetry is called simply laced $(|\alpha_1|=|\alpha_2|)$.
Otherwise, it is a non simply laced one
$(|\alpha_1|\neq|\alpha_2|)$.
\begin{table}[!ht]
\begin{center}
\begin{tabular}{|c|c|c|}
  \hline
  Lie Symmetry & $k_{12}$ & $k_{12}$ \\ \hline
   $A_1\bigoplus A_1 $  & $0$ & 0 \\ \hline
$A_2$ & -$1$& -1  \\ \hline
   $B_2$ & -$1$ & -2 \\ \hline
   $G_2$ & -$1$& -3 \\
  \hline
\end{tabular}
\end{center}
\label{tab1} \caption{Classification of rank two  Lie symmetries in
terms of the Cartan matrices}
\end{table}

\section{ Lie symmetries and  pure  2D  materials}

As mentioned in the introduction,   Lie symmetries has been explored
to study   the physical behaviors of the  hexagonal geometries
generalizing the structure  appearing in the graphene using
theoretical and experimental  methods.  In particular, an
experimental treatment on the hexagonal materials by  scanning
tunneling spectroscopy (STS) has been used  to   generate  a
$(\sqrt{3}\times \sqrt{3})R30^\circ$ super cell structure on  the
graphene like models, in contrast to the usual structure known by
(1$\times$1)\cite{8}.  In fact, the $(\sqrt{3}\times
\sqrt{3})R30^\circ$  structure  arises    naturally in the
construction of the $G_2$ root system  \cite{9,10}. Based on this
observation, a combination of $(\sqrt{3}\times \sqrt{3})R30^\circ$
and $(1\times 1)$ supercells has been proposed   to  model   new
materials  based on such  double  hexagonal Lie symmetries.
\cite{11}.

Inspired from these  activities, we propose a correspondence between
the root systems of rank two Lie symmetries and the geometry  of 2D
material physics. This may offer a novel  way to study  such
materials. Here, thought,  we will be concerned with similarities.
The link, that we are looking for,  can be supported by the
interplay between the hexagonal materials and the $A_2$   Lie
symmetries.

In order to see this point, let us first  consider the
 hexagonal materials having   only the (1$\times$1) structure. The
 latter appears in valence-four elements. Indeed,  there is a similarity  between  the    hexagonal  unit
 cell  of such materials  and
the root system of $A_2$ Lie symmetry. It is recalled that its
dimension reads as
\begin{equation}
 \mbox{Dim}\;A_2 = 2+6.
\end{equation}
The  corresponding   root system is  built from two simple roots
$\alpha_1$ and $\alpha_2$    of unequal length at $120^\circ$ angle
satisfying  following constraint
\begin{equation}
 \frac{|\alpha_2|^2}  {|\alpha_1|^2}=1.
\end{equation}
The  six nonzero roots come from these simple ones, with the sum and
the opposed,  forming  the well known hexagon ($A_2$-hexagon).  In
fact, the positions of the six atoms placed on  hexagonal  unit cell
are associated now with these six nonzero roots: $\{\pm \alpha_1,
\pm \alpha_2,  \pm (\alpha_1+\alpha_2)\}$.  An inspection shows that
one can elaborate a dictionary between the $A_2$ root system and the
hexagonal materials based on the  $(1\times 1)$ geometry.   This can
be formulated as follows:
\begin{itemize}
  \item  An   atom $A$  placed on  the hexagonal unit cell is associated with a
  $A_2$ nonzero root.
  \item  The lattice parameter $a$,  used in the hexagonal material,    corresponds  to  the length of the roots
  \begin{equation}
  a= |\alpha_1|=  |\alpha_2|.
\end{equation}
\item The general  material configuration  with  the  flat geometry can be
obtained by using the fact that  the $A_2$ hexagons   tessellate the
full plane forming the (1$\times$1) supercell structure.
\end{itemize}
\begin{center}
\begin{figure}[!ht]
\begin{center}
{\includegraphics[scale=0.7]{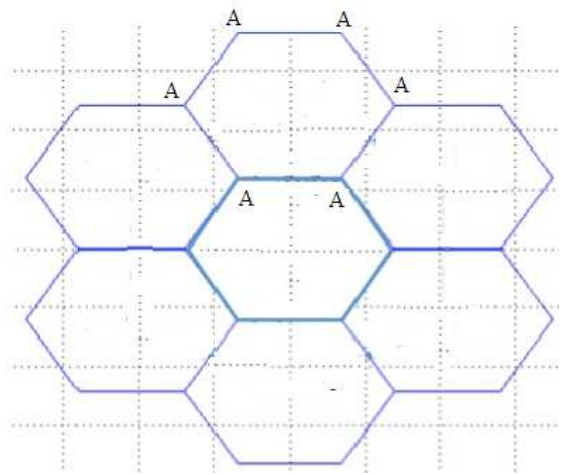}}
\end{center}
 \caption{$A_2$ hexagonal materials.} \vspace*{-.2cm}
\end{figure}
\end{center}

We naturally observe  that  the  $G_2$ Lie symmetry   can be also
incorporated in the discussion. This may be explored to   engineer
new $2D$ hexagonal materials. It is  worth  noting  that  $G_2$
 is an exceptional Lie symmetry with rank 2 and dimension
14. In this  way, eq(\ref{dim}) can be written as
\begin{equation}
 \mbox{Dim}\;G_2 = 2+12.
\end{equation}
This symmetry has been extensively studied in  the string
compactification in connection with a particular seven real
dimensional manifold, called $G_2$ manifold. It has been shown that
it can play a crucial role in the M-theory compactification
generating  semi-realistic   models with only four supercharges
 in our univers\cite{13,130,14}.  Roughly speaking, the root system  of  $G_2$
 symmetry contains  a special hexagonal root
structure residing on   two hexagons of unequal side lengths
generated by two simple unequal roots at angle
\begin{equation}
 (\widehat{\alpha_1, \alpha_2})=150^\circ= 120^\circ+30^\circ. \qquad
\end{equation}
These two simples roots satisfy  the constraint
\begin{equation}
\frac{|\alpha_2|^2}{ |\alpha_1|^2}=3,  \qquad
\end{equation}
leading to the following    Cartan matrix
 \begin{equation} \langle\alpha_i, \alpha_j\rangle=K_{ij}=\left(
  \begin{array}{cc}
    2 & -3 \\
    -1 & 2\\
  \end{array}
\right).
\end{equation}
This matrix can be encoded in   a  geometric  graph  called Dynkin
diagram. It is  recalled that,   the diagonal elements correspond to
two nodes and the non diagonal elements describe   the number of the
lines between them. In fact,  the  number of the lines between the
node $1$ and the node $2$ is given by   $k_{12}k_{21}$.  The Dynkin
diagram associated with  the  $G_2$  Lie symmetry  is presented in
figure (2). As known,  it is obtained from the one of the $D_4$
($so(8)$) Lie algebra as illustrated in figure 2.

\begin{center}
\begin{figure}[!ht]
\begin{center}
{\includegraphics[scale=0.7]{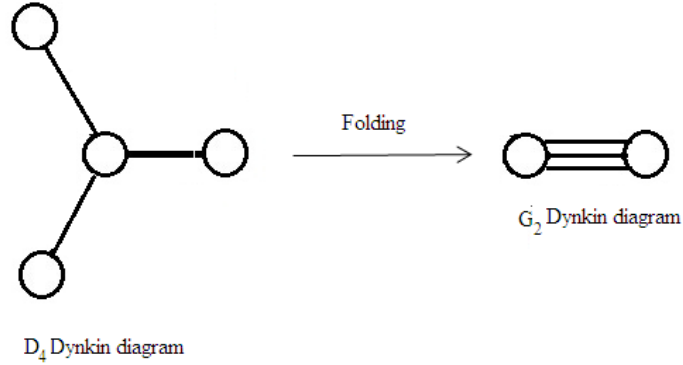}}
\end{center}
 \caption{$D_4$ and $G_2$ Dynkin diagrams.}  \vspace*{-.2cm}
\end{figure}
\end{center}

In fact,   each simple root of  the $G_2$ Lie symmetry generates a
single hexagon. The small one is defined  by the root set $\{ \pm
\alpha_1,\pm (\alpha_1+\alpha_2), \pm (2\alpha_1+\alpha_2) \}$ while
the second one is rotated by $30^{0}$ and  generated by $ \{ \pm
\alpha_2,\pm (3\alpha_1+\alpha_2), \pm(3\alpha_1+2\alpha_2) \}$. In
fact,  the $G_2$ hexagons  can be used to engineer materials with a
double honeycomb structure.  This involves two hexagons producing
materials having the property of being close to usual  one with one
periodic hexagon associated with the $A_2$ Lie symmetry
\cite{18,19}. It is evident  that the above correspondence  can be
extended  to $G_2$ Lie symmetry. It can be organized as follows:
\begin{itemize}
  \item  Each   atom placed on  the   double hexagonal unit
   cell, involving 12 atoms, is associated with a  root  of the  $G_2$ Lie
   symmetry.
  \item  The lattices parameters  $a_1$ and  $a_2$ of this unit cell
   correspond  to   the lengths of the two simple roots
  \begin{equation}
a_1=|\alpha_1|, \qquad a_2=|\alpha_2|=\sqrt{3}|\alpha_1|.
\end{equation}
\item The general  material structure with  the  flat geometry can be
obtained by using the fact that  the  $G_2$  hexagons   tessellate
the full plane forming a new  supercell crystal  structure  by
mixing the  $(1\times 1)$ and $(\sqrt{3}\times \sqrt{3})R30^\circ$
geometrical configurations. This geometry is illustrated  in  figure
3.
\end{itemize}

\begin{center}
\begin{figure}[!ht]
\begin{center}
{\includegraphics[scale=0.7]{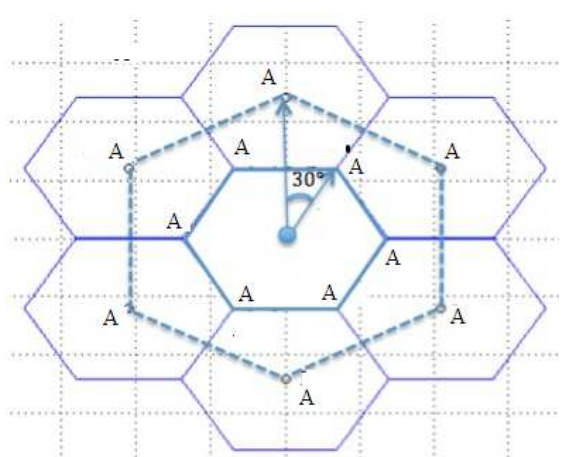}}
\end{center}
 \caption{$G_2$ hexagonal materials.}  \vspace*{-.2cm}
\end{figure}
\end{center}

Inspired from  the generalized Pythagorean theorem, the analysis
made for the  $G_2$  hexagons can be pushed  further. In particular,
we can do something similar  by  incorporating  the polyvalent
geometry
 generalizing  the trivalent  and
tetravalent geometry  appearing in so(8) Lie  symmetry. More
precisely,  we consider a particular geometry where a  central node
is connected with $3N$ legs as illustrated in figure 4. The
associated geometry  that we are looking for will be obtained by
using the standard techniques of folding of  the Dynkin nodes which
are permuted by an outer-automorphism group leaving the graph
invariant. Thus, starting form the polyvalent geometry and folding
the nodes that are permuted by $S_{3N}$ permutation group, we get
the graph corresponding to  the $G_2$  extended  Lie symmetry.  This
graph contains  two nodes connected by $3N$ lines as shown in figure
4.
\begin{center}
\begin{figure}[!ht]
\begin{center}
{\includegraphics[scale=0.7]{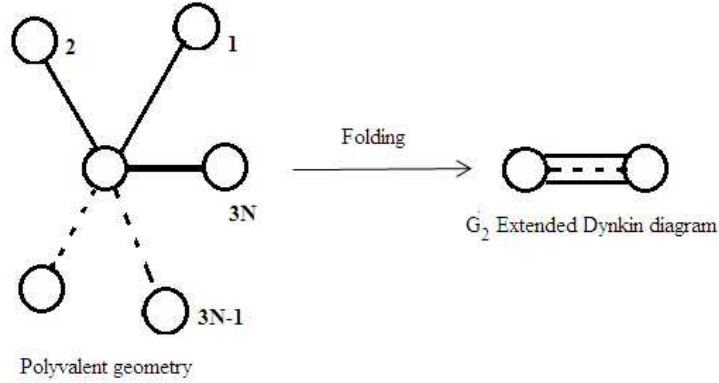}}
\end{center}
 \caption{$G_2$  extended  Dynkin diagram.} \vspace*{-.2cm}
\end{figure}
\end{center}

Using the relation  between the  Dynkin diagrams and  the Cartan
matrices, this  generalized version of the  $G_2$ Lie symmetry
should have the following  Cartan matrix
 \begin{equation}
 K_{ij}=\left(
  \begin{array}{cc}
    2 & -3N \\
    -1 & 2\\
  \end{array}
\right).
\end{equation}
This matrix should be  associated with   two simple roots
constrained by
\begin{equation}
 \frac{|\alpha_2|^2}  {|\alpha_1|^2}=3N.\qquad
\end{equation}
It is obvious from  this equation  that we can build up a double
hexagonal geometry based on the  generalized $G_2$ symmetry. It is
governed by the following points
\begin{itemize}
  \item  As in the case of $G_2$,  the   lattices parameters $a_1$ and  $a_2$   are linked  to
    the lengths of the simple
  roots. Correspondingly,  we  end up  with  the following  relations
  \begin{equation}
a_1=|\alpha_1|, \qquad a_2=|\alpha_2|=\sqrt{3N}|\alpha_1|
\end{equation}
\item The    supercell crystal  geometry   should  involve a
combined geometry  given   $(1\times 1)$ and $(\sqrt{3N}\times
\sqrt{3N})R30^\circ$ supercell structures.
\end{itemize}

To compte the discussion on rank two Lie symmetries, we incorporate
the remaining symmetries associated with   $B_{2}$ and $A_1\bigoplus
A_1 $. In fact,  the unit cell of the decorated square lattice
material can be linked  with the root system of  the $B_{2}$ Lie
symmetry generated by two simple roots and of unequal length at
$135$ angle. In this representation, each principal cell  is formed
by eight atoms. This double  geometry  contains two squares of
unequal side length at angle $45^{\circ }$. Each simple root of
 the $B_{2}$ symmetry generates a square. The small one is generated by
the root set $ \{\pm \alpha _{1},\pm (\alpha _{1}+\alpha _{2})\}$
describing a material  with only one square geometry. It is  worth
noting  that the geometry should be associated  with $A_1\bigoplus
A_1 $ Lie symmetry.   The big square is generated by the following
equal side length $\{\pm \alpha _{2},\pm (2\alpha _{1}+\alpha
_{2})\}$ describing  also a square geometry in material physics.

A close inspection reveals that there is a  dictionary  between two
rank Lie symmetries and  2D  material physics. This can be organized
in  Table 3.
\begin{table}[!ht]
\begin{center}
\begin{tabular}{|c|c|}
\hline Lie symmetries & Materials
 \\ \hline Root systems & Unit cells \\ \hline Non zero roots & Atom positions \\
 \hline $|\Delta|$ & Number of atoms in the material  unit cells
\\
 \hline
 Dimension of Cartan Sup-algebras & Dimension of material
 spaces\\
 \hline
 Simply laced Lie symmetries &  Materials  with  single  geometry \\
 \hline Non  simply laced Lie symmetries &  Materials  with
 double geometry
\\ \hline
\end{tabular}
\end{center}
\label{tab2} \caption{Correspondence between Lie symmetries and 2D
material physics.}
\end{table}

\section{Lie supersymmetries and doping material geometries}
On the basis of the above   correspondence, it is natural to think
about models  based on other Lie symmetries. In this section, we
consider the   Lie supersymmeries associated with fermionic degrees
of freedom. These  symmetries have been explored in many context in
physics  and can be thought as a possible extension of the one
discussed in the previous section \cite{15,16,17,18}. In fact, we
expect that Lie supersymmeries can be used to engineer  a new class
of materials. This may support the physical application of these
symmetries not only in string theory and related models but also in
nano-science technology.

In higher dimensional theories,   there are  many examples of such
Lie symmetries. The general study is beyond the scope of the present
work, though we will consider an  explicit example  A(1,0) which can
be thought of as a particular extension of  the $A_2$ Lie bosonic
symmetry.  This  symmetry  which has been developed  to study
aspects of supersymmetric integrable conformal models is quite
similar to $A_2$ but with some differences. It is an eight
dimensional Lie symmetry with rank 2. The corresponding root system
has been extensively studied involving fermmionic and bosonic roots.
In fact, it is shown that this symmetry involves two different root
systems. The first one has two fermionic nonzero roots $\alpha_1$
and $ \alpha_2 $ having a  length square zero, and a normalized
bosonic nonzero root $(\alpha_1+\alpha_2) $ with  a length square 2.
In this case, the total root system is given by $\{\pm\alpha_1,
\pm\alpha_2,\pm(\alpha_1+\alpha_1)\}$.  The second root system
involves  one  fermionic  simple nonzero root $\alpha_1$ having
length square zero and a simple   bosonic root $\alpha_2 $ with
length square 2.  The normalized fermionic one nonzero root with
length square 0 is  $(\alpha_2-\alpha_1) $. The total root system is
given by $\{\pm\alpha_1, \pm\alpha_2,\pm(\alpha_2-\alpha_1)\}$. It
has been shown that  both of them has a hexagonal geometry involving
fermionic and bosonic roots.  This indicates that  the theory of Lie
superalgebras differs from the theory of Lie symmetries.  This
difference in nature of roots drives us to think that such
symmetries could be associated  with magnetic   doping hexagonal
materials.  In the study graphene interacting with metal atoms, it
has been observed   that there are three adsorption sites named by
hollow H, bridge B and top T. In fact, the hollow H site is placed
at the center of a hexagon. In connection with Lie symmetries, these
sites should be associated with zero roots corresponding to  the
Cartan sub-algebras.  The bridge B site is located at the midpoint
of a carbon-carbon bond. Indeed, this can be  associated with the
length of simple roots via the the generalized Pythagorean theorem.
However, the top T site is placed directly above a carbon atom.
Considering atoms preferring the   T sites, the corresponding
structures could be interpreted as geometries of  the  root system.
This is motivated by the fact  that in the hexagonal unit cell
material contains atoms with different nature. In this way, the
metal atom positions could be associated with fermionic roots, while
the ordinary atoms are associated with the  bosonic roots. This
observation could be explored further to give a complete picture of
the dictionary  presented in Table 3.

\section{Speculations from string theory}
In this paper, we  have developed a new method to  approach to  2D
material physics. In particular, we have presented  a dictionary
between rank two  Lie symmetries   and the geometry of 2D  material
physics. In this way,  the material unit cell   has  been
interpreted as a root system. More precisely, we have analyzed in
some details the $A_2$ and $G_2$ Lie symmetries   and   found they
are linked to hexagonal materials. This approach could  be applied
to  Lie supersymmetries corresponding to fermionic  degrees of
freedom providing   a new way to investigate doping material
geometries. We intend  to investigated elsewhere other symmetries to
complete  the Table 3 by introducing other Lie symmetries used in
modern  physics.

This work comes up with  the opening of many windows.  We would like
to stress  that a possible connection with  string theory can be
given. In fact, the correspondence presented here  shears  many
similarities with the link
  between  singularity theory  and Lie symmetries.
 This link  based on a  correspondence between the ADE roots  and
 the two-cycles used in in the deformation of ADE singularities of ALE spaces considered as local versions
 of the K3 surface. In fact, to each
simple root one  associates  a   complex projective space
$\mathbb{CP}^1$ used in the blowing up of singular points. This
connection has been considered as an important development in the
string theory compactification  supported by the string-string
duality  in six dimensions and  D-brane configurations on which
 gauge theories live. This has been  controlled by the root systems including the hexagonal one explored
  in particle physics\cite{180}.

Inspired from these connections, one can  speculate on string theory
interpretation  in  terms of  $(p,q)$ string junctions associated
with root systems of simply and non simply laved Lie symmetries
developed in  \cite{19,20}. In this picture, we anticipate that the
material physics can be studied using brane physics. It is  recalled
that a brane is an extended object upon which open strings can end
with appropriate boundary conditions.  In fact, the $A_2$ material
unit cell can be represented by     open strings stretched on branes
with similar behaviors. Indeed, writing the two simple roots in
terms of a real basis $\{e_i,\; i=1,2,3\}$ of the orthonormal
vectors of $R^3$
 \begin{eqnarray}
\alpha_i=e_i-e_{i+1}, \qquad i=1,2
\end{eqnarray}
 it is possible to give an interpretation  in terms of three branes
associated with  the basis $\{e_i\}$.  The corresponding
$(p_i,q_i)$ brane charge webs should satisfy the vanishing charge
condition
 \begin{eqnarray}
 \label{charge}
p_1+p_2+p_3&=&0\\ \nonumber q_1+q_2+q_3&=&0
\end{eqnarray}
required by   the $A_2$  root system.
 The total  six  $(p_i,q_i)$  open strings
can be  obtained  from  the  mirror branes with   the  opposed
charges. A possible charge vector can given by solving the above
equations. We expect that the corresponding  physics content  can be
determined explicitly from the geometry of the $(p, q)$ networks.
This formulation is quite similar to toric realization of the
$\mathbb{CP}^2$ projective space \cite{21}. It   is a complex two
dimensional manifold having an $\mbox{U(1)}^2$ toric action
exhibiting three fixed points associated with an equation similar to
(\ref{charge}). It would therefore be of interest to try to extract
information on material physics  based on  such mathematical
backgrounds. We believe that these questions deserve to be
investigated further.

\end{document}